\documentclass[prd,preprintnumbers,twocolumn,eqsecnum,floatfix,a4paper,nofootinbib,superscriptaddress]{revtex4-2}

\usepackage{color}
\usepackage[dvipsnames]{xcolor}
\usepackage{amsmath,amssymb,graphicx,mathtools}
\usepackage{bm,lipsum}
\usepackage{times}
\usepackage{microtype}
\usepackage[utf8]{inputenc}
\usepackage{booktabs}
\usepackage[normalem]{ulem}
\usepackage[varg]{txfonts}
\usepackage{bm,graphics,graphicx,epsfig,xcolor,ulem}
\usepackage[colorlinks, pdfborder={0 0 0}]{hyperref}
\usepackage{multirow}
\usepackage{pifont}
\usepackage{array,threeparttable}
\usepackage{diagbox}
\usepackage{boldline}
\usepackage{gensymb}
\usepackage{url}
\usepackage{acronym}
\usepackage[capitalise]{cleveref}

\definecolor{LinkColor}{rgb}{0.75, 0, 0}
\definecolor{CiteColor}{rgb}{0, 0.5, 0.5}
\definecolor{UrlColor}{rgb}{0, 0, 0.75}
\hypersetup{linkcolor=LinkColor}
\hypersetup{citecolor=CiteColor}
\hypersetup{urlcolor=UrlColor}
\allowdisplaybreaks
\newcommand{\arnab}[1]{\textcolor{MidnightBlue}{\textbf{#1}}}

\begin{document}
\title{Remnant properties of binary neutron star mergers undergoing prompt collapse}


\newcommand{\psuigc}{\affiliation{Institute for Gravitation and the Cosmos, The Pennsylvania State University, University Park, PA, 16802, USA}}
\newcommand{\psuphys}{\affiliation{Department of Physics, The Pennsylvania State University, University Park, PA, 16802, USA}}
\newcommand{\psuastro}{\affiliation{Department of Astronomy \& Astrophysics, The Pennsylvania State University, University Park, PA, 16802, USA}}
\newcommand{\aei}{\affiliation{Max Planck Institute for Gravitational Physics (Albert Einstein Institute), Am Mühlenberg 1, Potsdam 14476, Germany}}
\newcommand{\cardiff}{\affiliation{School of Physics and Astronomy, Cardiff University, Cardiff, UK, CF24 3AA}}
\newcommand{\infn}{\affiliation{Dipartimento di Fisica, Università di Trento, Via Sommarive 14, 38123 Trento, Italy}}
\newcommand{\trento}{\affiliation{INFN-TIFPA, Trento Institute for Fundamental Physics and Applications, Via Sommarive 14, I-38123 Trento, Italy}}
\newcommand{\UPisa}{\affiliation{Dipartimento di Fisica, Universit\`{a} di Pisa, Largo B.  Pontecorvo, 3 I-56127 Pisa, Italy}}
\newcommand{\infnPisa}{\affiliation{INFN, Sezione di Pisa, Largo B. Pontecorvo, 3 I-56127 Pisa, Italy}}
\newcommand{\iitb}{\affiliation{Department of Physics, Indian Institute of Technology Bombay, Mumbai 400076, India}}
\newcommand{\fbk}{\affiliation{Data Science for Industry and Physics, Fondazione Bruno Kessler, via Sommarive 18, 38123, Trento (TN), Italy}}

\author{Arnab Dhani}
\email{arnab.dhani@aei.mpg.de}
\aei \psuigc \psuphys

\author{Alessandro Camilletti}
\infn \trento \fbk

\author{David Radice}
\thanks{Alfred P. Sloan fellow}
\psuigc \psuphys \psuastro

\author{Rahul Kashyap}
\iitb \psuigc \psuphys

\author{Bangalore Sathyaprakash}
\psuigc \psuphys \cardiff

\author{Domenico Logoteta}
\UPisa \infnPisa

\author{Albino Perego}
\infn \trento

\begin{abstract}
We study the properties of remnants formed in prompt-collapse binary neutron star mergers. We consider non-spinning binaries over a range of total masses and mass ratios across a set of 22 equations of state, totaling 107 numerical relativity simulations. We report the final mass and spin of the systems (including the accretion disk and ejecta) to be constrained in a narrow range, regardless of the binary configuration and matter effects. This sets them apart from binary black-hole merger remnants. We assess the detectability of the postmerger signal in a future 40 km Cosmic Explorer observatory and find that the signal-to-noise ratio in the postmerger of an optimally located and oriented binary at a distance of 100 Mpc can range from ${<}1$ to 8, depending on the binary configuration and equation of state, with a majority of them greater than 4 in the set of simulations that we consider. We also consider the distinguishability between prompt-collapse binary neutron star and binary black hole mergers with the same masses and spins. We find that Cosmic Explorer will be able to distinguish such systems primarily via the measurement of tidal effects in the late inspiral. Neutron star binaries with \emph{reduced tidal deformability} $\tilde\Lambda$ as small as ${\sim}3.5$ can be identified up to a distance of 100 Mpc, while neutron star binaries with $\tilde\Lambda\sim22$ can be identified to distances greater than 250 Mpc. This is larger than the distance up to which the postmerger will be visible. Finally, we discuss the possible implications of our findings for the equation of state of neutron stars from the gravitational-wave event GW230529.
\end{abstract}

\maketitle

\section{Introduction}
\label{sec:introduction}
The observation of gravitational waves from compact binary mergers \cite{LIGOScientific:2016aoc, KAGRA:2021vkt} have informed us of the astrophysical population of such sources, allowed for precision tests of strong-field gravity \cite{LIGOScientific:2021sio, LIGOScientific:2018dkp}, provided an independent measurement of the expansionary history of the Universe \cite{LIGOScientific:2017adf, LIGOScientific:2021aug}, and offered a tantalizing prospect of inferring the \ac{NS} \ac{EoS} \cite{LIGOScientific:2019eut}. With improving sensitivities of ground-based detectors \cite{LIGOScientific:2014pky,VIRGO:2014yos,KAGRA:2018plz} and the ongoing planning for future observatories \cite{Reitze:2019iox,Punturo:2010zza,LIGO:2020xsf}, the precision with which we can measure source properties \cite{Borhanian:2022czq,Iacovelli:2022bbs,Abac:2025saz} and infer the nature of gravity in the strong-field regime and on large scales will improve many fold. Specifically, the proposed next-generation (XG) of ground-based gravitational-wave detectors, such as \ac{CE}~\cite{Reitze:2019iox} and Einstein Telescope~\cite{Punturo:2010zz}, can observe $O({\rm few}\ M_\odot)$ compact binary mergers up to a redshift of $z\sim1$ \cite{Borhanian:2022czq} and measure the radius of NS at a precision of a few tens to 100 m \cite{Evans:2021gyd,Gupta:2023lga,Branchesi:2023mws,Evans:2023euw,Abac:2025saz,Khan:2015jqa,Huxford:2023qne}. 

This population of sources can be comprised of \acp{BNS}, \acp{BBH}, or neutron-star black holes. It is imperative that one has an unbiased inference of the source class and its characteristics. For instance, LIGO-Virgo-KAGRA (LVK) observations alone cannot rule out the possibility that GW170817 \cite{LIGOScientific:2017vwq}, the first \ac{BNS} merger observed by the LVK Collaboration, and GW190425 \cite{LIGOScientific:2020aai} are low-mass \ac{BBH} mergers \cite{LIGOScientific:2019eut}. In the case of GW170817, it was the subsequent detection of short gamma ray burst and kilonova \cite{LIGOScientific:2017ync} 
that conclusively classified the event as an \ac{NS} merger. The absence of an electromagnetic (EM) counterpart for GW190425 means that the event is classified as a \ac{BNS} merger only because of its component masses \cite{LIGOScientific:2020aai}. However, under the assumption that the event was a \ac{BNS} merger, there is a $96\%-97\%$ probability that the remnant promptly collapsed to a \ac{BH} \cite{LIGOScientific:2020aai}.
An accurate source classification is of immense value for both astrophysical and fundamental physics inferences. First, one of the most important observables of interest in nuclear physics is the maximum mass of NSs, because it depends on the EoS at extreme densities and reveals the potential for NS cores to host hyperons or quarks \cite{Lattimer:2012nd}. Second, one of the widely used methods to test the nature of strong gravity and the dynamics of black hole binaries is the \emph{inspiral-merger-ringdown} consistency test \cite{Hughes:2004vw,Ghosh:2016qgn}. Therefore, an incorrect identification of a \ac{BNS} merger as a \ac{BBH} merger can lead to a false violation of general relativity~\cite{Gupta:2024gun}.

The outcome of a \ac{BNS} merger can be one of the following: a stable differentially rotating \ac{NS}, an unstable short-lived \ac{NS} undergoing a delayed collapse due to the inability of the differential rotation to support the remnant mass on a secular timescale, or a prompt collapse to a \ac{BH} on a dynamical timescale. The radiated energy and angular momentum of the system depend on the nature of the formed remnant. Long-lived NS remnants are efficient emitters of GWs, while, in a prompt collapse, the remnant---including the disk and ejecta---is expected to retain most of the mass and angular momentum of the system, 
even though the peak luminosity is greater in the latter case \cite{Zappa:2017xba}. The delineation of a prompt collapse and a remnant NS is given by the threshold mass $M_{thres}$ which is defined as the minimum total mass of a binary that leads to a prompt collapse \cite{Shibata:2005ss} and depends weakly on its mass ratio for close to equal mass systems \cite{Shibata:2005ss, Perego:2021mkd}. 

Prompt collapse mergers have been widely studied with a focus on understanding the properties of the threshold mass \cite{Shibata:2005ss, Hotokezaka:2011dh, Bauswein:2013jpa, Zappa:2017xba, Agathos:2019sah, Koppel:2019pys, Bauswein:2020aag, Bauswein:2020xlt, Perego:2021mkd, Kashyap:2021wzs, Kolsch:2021lub, Cokluk:2023xio} and its detectability \cite{Agathos:2019sah}, since its determination can inform the maximum mass and compactness supported by the EoS of NSs. 
This is because it is currently unknown whether core-collapse supernovae, expected to be the primary formation mechanism of NS in binaries, can produce NS across the entire mass range supported by a given EoS since most galactic NS lie between $1.3-1.4 \, M_{\odot}$ \cite{Ozel:2016oaf}.
However, the properties of the remnants, their astrophysical significance, and the detectability of generic prompt collapses with future detectors have only recently started getting attention~\cite{Kolsch:2021lub,Camilletti:2022jms,Dudi:2021abi,Dhani:2023ijt,Cokluk:2023xio}.

In the absence of an EM counterpart, \ac{BNS} mergers are distinguished from \ac{BBH} mergers by the presence of tidal interactions between the NSs in the former. However, heavier NSs are more compact and have smaller tidal deformabilities. Therefore, even if such heavy NSs are produced astrophysically and form binaries, it is currently unknown whether even future detectors can measure their tidal deformabilities with sufficient accuracy to confidently classify them as a \ac{BNS} merger. Consequently, it is currently not known what the maximum NS mass is that can be confidently measured using GW observations. These heavy binaries would also promptly collapse to a BH. 

In this paper, we use data from 107 nonspinning \ac{BNS} merger simulations, spanning 22 EoS and a wide range of masses and mass ratios, to investigate the properties of a GW signal from prompt-collapse \ac{BNS} mergers. We find that the predicted remnant mass and spin (including the disk) are significantly greater than those produced by a corresponding \ac{BBH} merger. 
Subsequently, we explore the detectability of the ringdown. We find that for a majority of systems, the ringdown can be observed up to a distance of 100 Mpc in a future observatory. Furthermore, we show the contrast in the signal power during the ringdown phase of a prompt collapse and its \ac{BBH} equivalent as observed in a 40 km Cosmic Explorer detector with the former being more silent. Additionally, we ask and take the first steps towards answering the following dual question: 
What is the smallest tidal deformability that can be distinguished from zero using the inspiral measurement?
Is it possible to distinguish between \ac{BNS} and \ac{BBH} mergers in the event that their inspiral signals are consistent with each other? 
We find that a future CE detector can distinguish virtually all \ac{BNS} and \ac{BBH} mergers up to a distance of 100 Mpc from the inspiral and a majority of these systems from their ringdowns too. At larger distances, the postmerger becomes very weak while the tidal deformabilities can still be well measured, depending on the stiffness of the EoS. Finally, we discuss the astrophysical consequences of identifying a prompt collapse with the progenitor of GW230529's primary component. 

The rest of the paper is organised as follows. In Sec. \ref{sec:nr_simulations}, we describe numerical relativity data used in this study. Following that, we report our results in Sec. \ref{sec:results} where we compare the remnant mass and spin of prompt collapse to corresponding \ac{BBH} mergers. Thereafter, we evaluate their detectability and distinguishability with \ac{BBH} merger ringdowns. We conclude in Sec. \ref{sec:conclusion} with a discussion of astrophysical implications and future work.

\section{Numerical relativity data}
\label{sec:nr_simulations}
\begin{figure}
    \centering
    \includegraphics[width=\columnwidth]{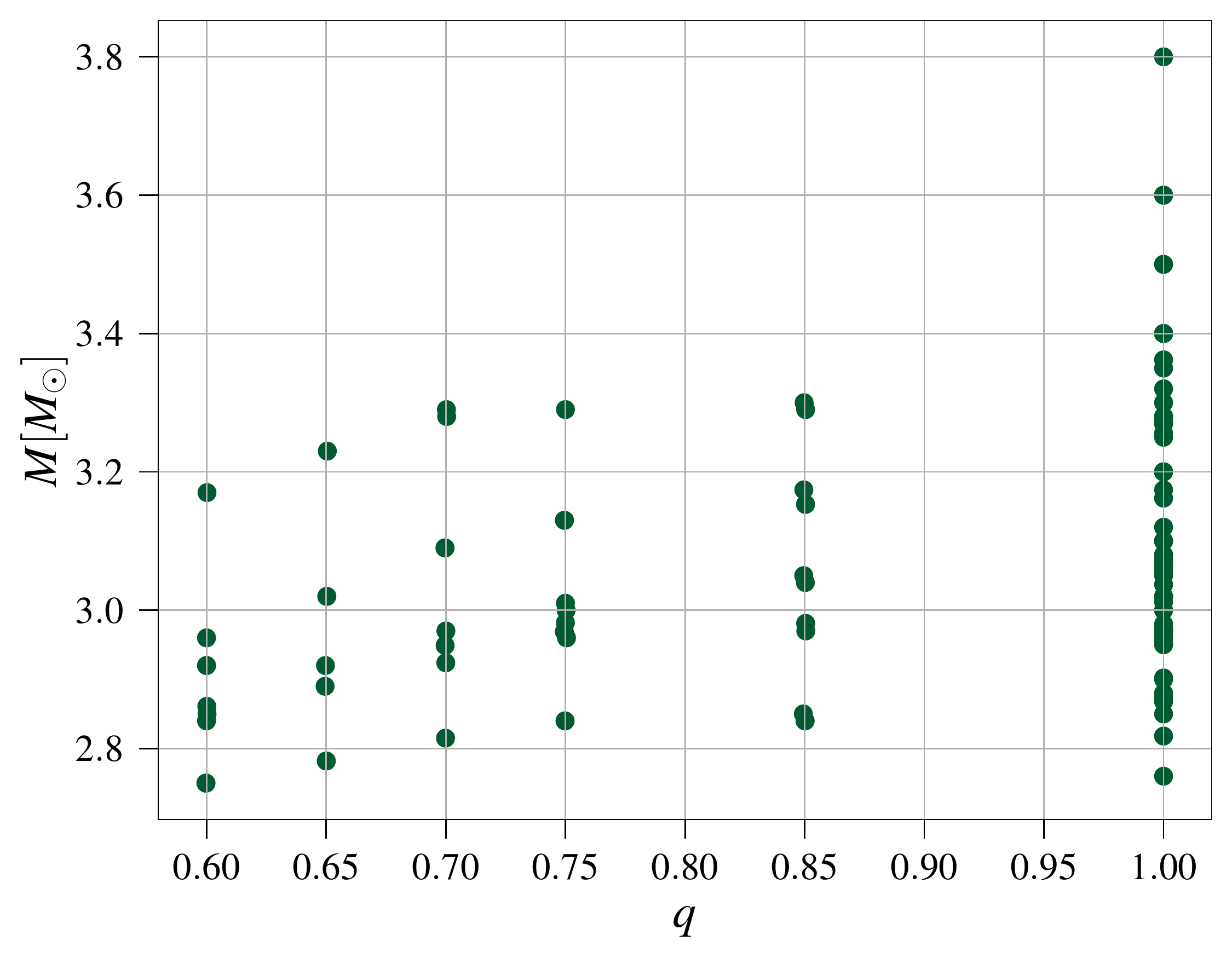}
    \caption{The total mass of the binary at infinite separation and mass ratio of the 107 simulations used in this study.}
    \label{fig:mtot_q}
\end{figure}
We consider 22 EoS spanning a wide range in the corresponding non-rotating NS maximum masses and typical radii. In particular, we employ 15 finite-temperature EoS: BHB$\Lambda\phi$ \cite{Banik:2014qja}, BLh \cite{Bombaci:2018ksa,Logoteta:2020yxf}, DD2qG \cite{Logoteta:2022hvb}, 
HS (DD2) \cite{Typel:2009sy, Hempel:2009mc}, LS220 \cite{Lattimer:1991nc}, SFHo \cite{Steiner:2012rk}, SRO (SLy) \cite{Douchin:2001sv, Schneider:2017tfi}, and 9 variants of the SRO EOS with different values of the empirical nuclear parameters \cite{Schneider:2017tfi, Schneider:2019shi}. We also consider 6 zero-temperature $\beta$-equilibrated EoS: ALF2 \cite{Alford:2004pf}, H3 and H4 \cite{Glendenning:1991es, Lackey:2005tk, Read:2008iy}, the Big Apple EoS (BA from now on) \cite{Fattoyev:2020cws}, and the GRW1 and GRW2 EoSs \cite{Kashyap:2021wzs}. The zero-temperature EoSs are supplemented with thermal effects using an adiabatic index, $\Gamma_{th} = 1.7$ following, e.g., Refs.~\cite{Shibata:2005ss, Bauswein2010-ko, Endrizzi2018-np, Figura:2020ab}. GRW1 and GRW2 are completely phenomenological EoSs, while the other models are nuclear-theory based. Three EoSs include hyperons in addition to nucleons: BHB$\Lambda\phi$,
H3 and H4. Three EoSs include a QCD phase transition to deconfined quarks: ALF2, BLQ, and DD2qG.

We consider a total of 107 numerical relativity simulations. The distribution of total masses at infinite separation, $M$, and mass ratios, $q$, considered in this study is shown in Fig.~\ref{fig:mtot_q}. These simulations have already been presented in Refs.~\cite{Kashyap:2021wzs, Perego:2021mkd}, to which we refer for the details. See also \cite{Gonzalez:2022mgo} for an up to date description of the simulation code capabilities. In brief, all simulations were initialized using quasi-circular, irrotational binary initial data with the pseudospectral code \texttt{Lorene} \cite{Gourgoulhon:2000nn} and evolved with the  \texttt{WhiskyTHC} code \cite{Radice:2012cu, Radice:2013hxh, Radice:2013xpa, Radice:2015nva}. The latter is built on top of the \texttt{Einstein Toolkit} \cite{EinsteinToolkit:2022_11} and makes use of the \texttt{Carpet} adaptive mesh refinement (AMR) framework \cite{Schnetter:2003rb, Reisswig:2012nc}, which implements the Berger-Oliger scheme with refluxing \cite{Berger1984-uz, Berger1989-je}.
As in these previous works, we classify as a prompt collapse those binaries for which the lapse function decreases monotonically below 0.2 until BH formation (no bounce).

\section{Results}
\label{sec:results}
\subsection{Radiated energy and angular momentum}
\label{subsec:erad_jrad}
\begin{figure}
    \centering
    \includegraphics[width=\columnwidth]{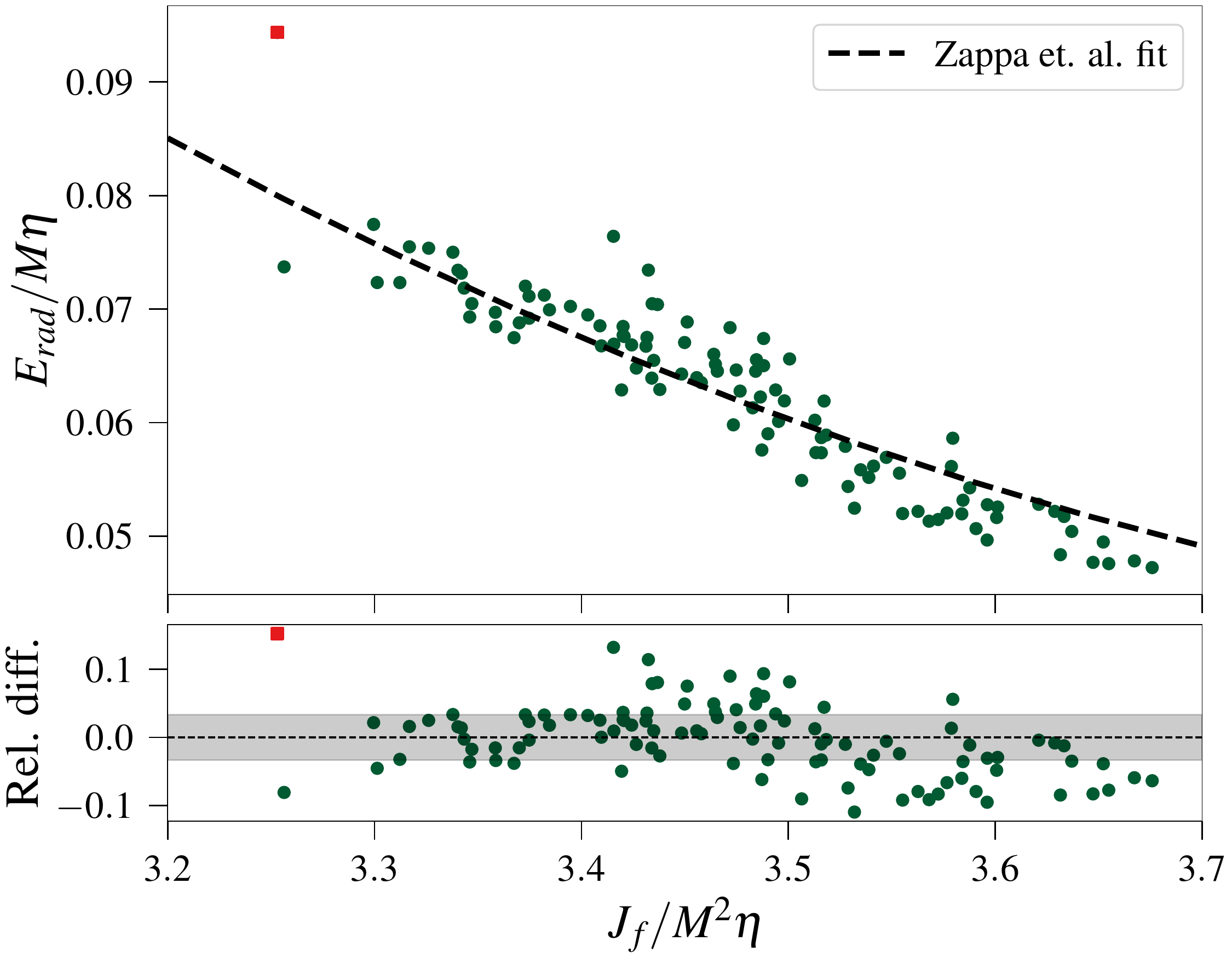}
    \caption{Reduced total radiated energy versus reduced angular momentum of the remnant. The fit given in \textcite{Zappa:2017xba} is overlayed in the top panel. The bottom panel shows the relative difference with respect to the fit and the gray band shows the median of the absolute relative difference. The outlier denoted with a red square is the $1.7\ M_\odot - 1.7\ M_\odot$ with the BA EoS, see the main text for further discussion.}
    \label{fig:erad_jrem}
\end{figure}

Energy and angular momentum are radiated during a compact binary merger in the form of gravitational waves, with the total radiated energy and angular momentum denoted by $E_{rad}$ and $J_{rad}$. The former is calculated from the News tensor while the latter also uses the strain as input \cite{Damour:2011fu}. In practice, the News tensor is calculated by differentiating the strain. \textcite{Zappa:2017xba} showed that the angular momentum of the merged remnant is uniquely determined by the total radiated energy. It was reported that this relationship holds regardless of the binary parameters, the EoS employed, or the microphysics incorporated in the simulations during the postmerger. 

In Fig.~\ref{fig:erad_jrem}, we plot the reduced radiated energy as a function of the reduced remnant angular momentum for the set of simulations used in this study. We also depict the quadratic fit given in \citet{Zappa:2017xba} for these quantities with the bottom panel showing the residuals.
We found that the coefficients of the fit reported in \citet{Zappa:2017xba} do not have sufficient significant digits required to reproduce the curve, and also contain a typographic error. The correct fit is given by [F.~Zappa private communication],
\begin{equation}
    e_{rad} = c_2 j_{rem}^2 + c_1 j_{rem} + c_0
\end{equation}
where $e_{rad}=E_{rad}/(M\eta)$ and $j_{rem}=J_f/(M^2\eta)$; $\eta$ is the symmetric mass ratio and $J_f$ is the angular momentum of the remnant. The coefficients of the fit are $c_2=0.05334342$, $c_1=-0.43994502$, and $c_0=0.94665103$. We find a median relative difference of ${\sim}3.3\%$ and a maximum of ${\sim}15\%$. 

The binary with the largest deviation---residing at $3\sigma$ in the residual distribution and marked by a red square---in \cref{fig:erad_jrem} is the $1.7\ M_\odot - 1.7\ M_\odot$ with the BA EoS. This is a high mass system that produces a short-lived (${\sim}2$~ms) remnant that collapses to a BH without a bounce (so it is classified as prompt collapse even though BH formation is not ``immediate''). The BA EoS is derived from a relativistic mean-field theory with parameters tweaked to produce an extremely large NS maximum mass while remaining consistent with current astronomical and laboratory constraints \cite{Fattoyev:2020cws}. The $1.7\ M_\odot - 1.7\ M_\odot$ binary is just above the threshold mass $M_{thr} \simeq 3.38\ M_\odot$ for this EoS \cite{Kashyap:2021wzs}. The proximity to the threshold mass and the extreme properties of the BA EoS are likely the reason why this binary is an outlier in Fig.~\ref{fig:erad_jrem} and in subsequent figures, where it is marked with a red square.

The existence of such a relation for \ac{BNS} mergers is significant because it shows that the angular momentum of the remnant is primarily governed by the gravitational dynamics and is insensitive to the details of the hydrodynamics of these systems. Since the systems considered here are prompt collapses, the post-merger timescale is at most a few milliseconds, following which the remnant settles down into an isolated Kerr BH. As noted in \cite{Zappa:2017xba}, this relation can be used to calculate the final spin of the remnant BH, especially with next-generation detectors which can measure the total radiated energy. Next-generation detectors may also be able to detect the quasi-normal mode ringing of the remnant BH and measure the QNM frequency, providing an independent estimate of the final mass and spin. However, one needs to be cognizant that for unequal mass mergers, significant mass and angular momentum is contained in the accretion disk around the remnant \ac{BH}~\cite{Dhani:2025xno,Kolsch:2021lub,Camilletti:2022jms,Dudi:2021abi} and needs to be taken into account. A cross-validation of both measurements would be a crucial check of the dynamics near the merger when the gravitational radiation is maximum.

\subsection{Remnant mass and spin}
\label{subsec:remnant_mass_spin}
\begin{figure*}
    \centering
    \includegraphics[width=2\columnwidth]{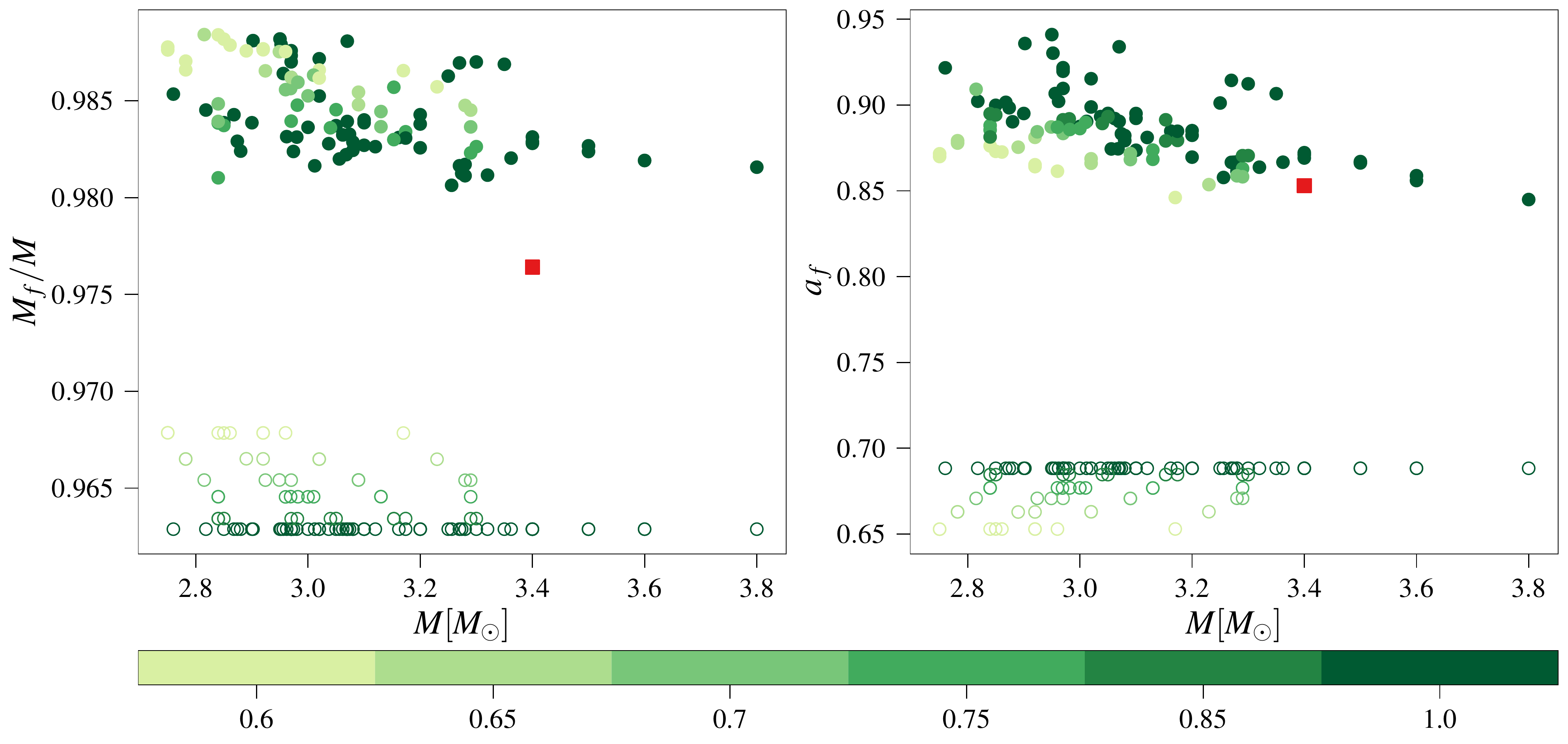}
    \caption{Dimensionless final mass (left) and final spin (right) as a function of the total mass of the binary. Unfilled markers show the final mass and spin if the system was a binary black hole. The color gradient shows the asymmetry of the system with lighter shades representing more asymmetric binaries. \arnab{Note that the colorbar is not uniform since simulations are only available at these fixed mass ratio values.} The outlier denoted with a red square is the $1.7\ M_\odot - 1.7\ M_\odot$ with the BA EoS, see the main text for further discussion.}
    \label{fig:mf_af}
\end{figure*}

A Kerr black hole is formed from the prompt collapse of a quasi-circular binary neutron star merger. For equal mass mergers, there is negligible accretion of matter around the collapsed BH. However, for unequal mass mergers, significant mass and angular momentum is contained in the accretion disk around the remnant \ac{BH}~\cite{Dhani:2025xno,Kolsch:2021lub,Camilletti:2022jms,Dudi:2021abi}. The final mass $M_f$ and dimensionless spin $a_f$ of the remnant are estimated by (including the mass and angular momentum in the ejecta/disk),
\begin{equation}
    M_f = M_{adm} - E_{rad},
\end{equation}
\begin{equation}
    a_f = \frac{J_{adm} - J_{rad}}{M_f^2},
\end{equation}
where $M_{adm}$ and $J_{adm}$ are the ADM mass and angular momentum of the system at the beginning of the simulations. 
Note that the ADM mass differs from the total mass in that it also contains the energy of the gravitational field.
For a binary system, the gravitational field energy includes the binding energy and the energy in gravitational waves. 

The mass (left panel) and dimensionless spin (right panel) of the remnant (including the disk) for the set of binary configurations and EoSs considered in this study are shown in Fig.~\ref{fig:mf_af} with filled markers. Concurrently, we also show the remnant properties for non-spinning \ac{BBH} binaries having the same total mass and mass ratio as the systems under consideration with unfilled markers. The final mass and dimensionless spin in this case is calculated using the fitting formulas for initially non-spinning \ac{BBH} given in \textcite{Berti:2007fi}. The asymmetry of a system is depicted using a color gradient with lighter shades of green representing more asymmetric binaries. 

It is immediately noticeable from the figure that the remnant mass and dimensionless spin of the promptly collapsing systems are larger than the corresponding black hole binaries, signifying that less energy and angular momentum are radiated during the merger in the former compared to the latter. Where, for a binary black hole merger, 3-4\% of the total initial mass of the binary is radiated away in the form of gravitational waves, for a prompt collapse, only 1-2\% of the initial total mass is radiated. Note that this is contrary to the case of a \ac{BNS} merger forming a NS remnant where the radiated energy is comparable to, though always less than, an equivalent \ac{BBH} merger \cite{Bernuzzi:2015opx}. Moreover, for a \ac{BBH} merger, the total mass of the system is a scaling factor and the scaled final mass is a function of only the mass ratio, with a more asymmetric system radiating less energy. For a \ac{BNS} merger, the total mass is not just a scaling factor as is clear by the spread in $M_f/M$ in Fig.~\ref{fig:mf_af}. However, for a prompt collapse, this spread is very narrow and less than 1\%. Due to this spread, there is no clear dependence on the mass ratio even though the overall trend is that asymmetric systems radiate less for most cases. 

The dimensionless final spin of an equal mass, initially non-spinning \ac{BBH} system is $\approx0.69$ with the final spin decreasing for more asymmetric systems (see right panel of Fig.~\ref{fig:mf_af}). Again, the final spin depends only on the mass ratio of the system and not on the initial total mass. For the prompt collapse systems, the smallest value of the dimensionless spin we find is $a_f\sim0.85$, with the largest value reaching $a_f\sim0.95$. Moreover, there is a spread in the values of the dimensionless final spin, which depends on the total mass of the system and the EoS. However, as in the case of the scaled final mass, the spread is narrow. Furthermore, one can qualitatively see that more asymmetric systems have lower final spin.

\begin{figure}
    \centering
    \includegraphics[width=\columnwidth]{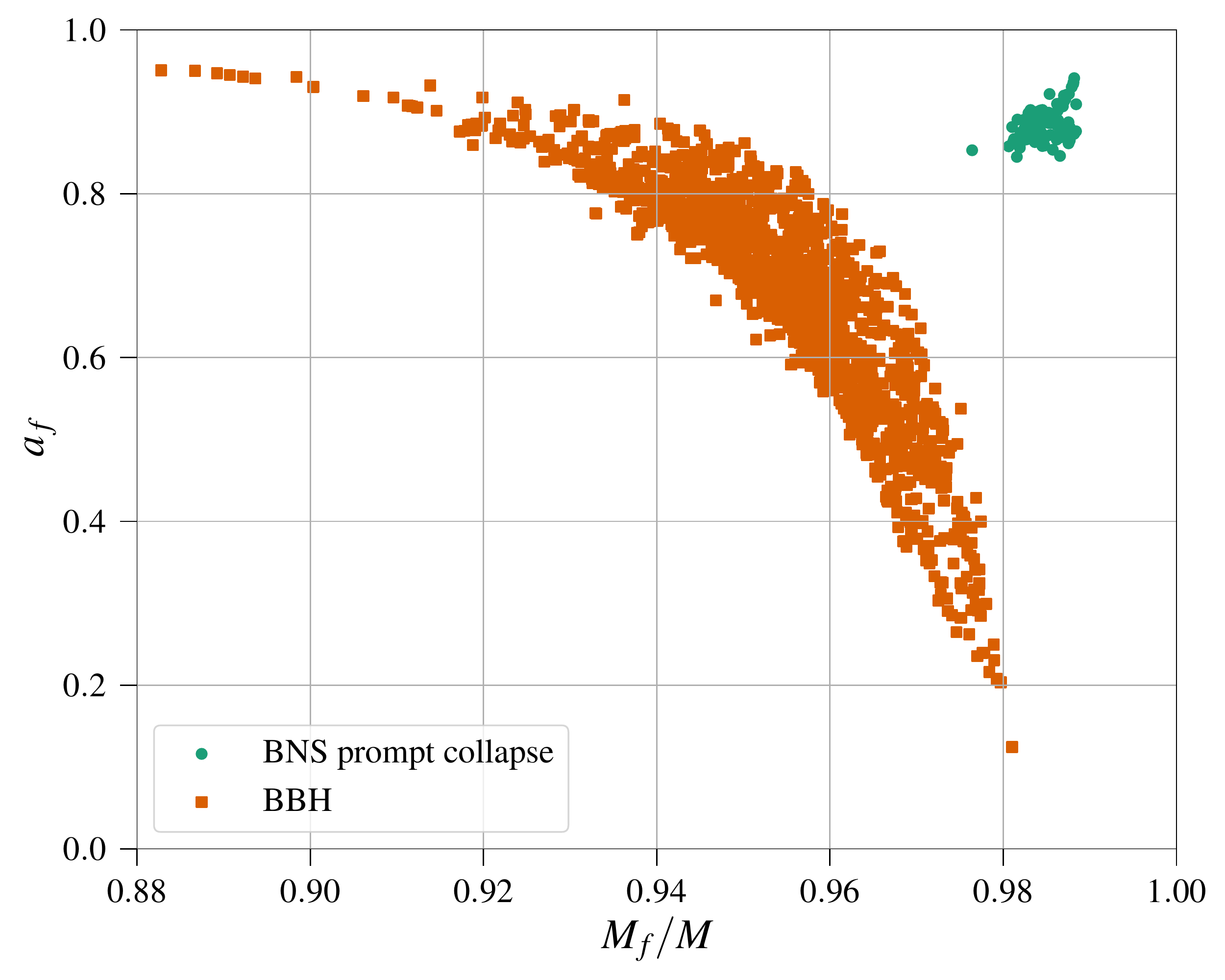}
    \caption{The remnant mass and spin of \ac{BBH} mergers versus \ac{BNS} prompt collapses. There are 1937 \ac{BBH} mergers in this figure corresponding to all the publicly available SXS simulations with $q \leq 3$ for which such information is available. The outlier denoted with a red square is the $1.7\ M_\odot - 1.7\ M_\odot$ with the BA EoS, see the main text for further discussion.}
    \label{fig:BBH_BNS_mf_af}
\end{figure}

Motivated by the observation that the remnant mass and spin of prompt collapse mergers (including the disk) differ from the equivalent \ac{BBH} system, one can ask which BBH configurations produce analogous remnants.
To probe this, we consider the catalog of publicly available noneccentric SXS simulations~\cite{Boyle:2019kee,Scheel:2025jct} of \acp{BBH} up to a mass ratio of 3, since a \ac{BNS} system is not expected to be more asymmetric. This gives us a set of 
1937
\ac{BBH} simulations containing non-spinning systems, aligned spin systems, and precessing systems. 
We plot them in the remnant mass -- spin parameter space together with all the prompt collapse mergers used in this study in Fig.~\ref{fig:BBH_BNS_mf_af}. Astonishingly, we find that the two classes of sources lie in entirely disjoint areas of the parameter space. There is a tight negative correlation for \ac{BBH} systems where the production of highly spinning remnants necessitates the loss of a large amount of energy through gravitational radiation. This is directly opposite of the trend seen in Fig.~\ref{fig:erad_jrem} where a smaller radiated energy means a larger remnant angular momentum. In fact, we find that the smallest scaled angular momentum ($J_f/M^2=M_f^2a_f/M^2$) of a prompt collapse \ac{BNS} remnant is greater than the largest value across all the \ac{BBH} simulations considered (see \cref{sec:appA}).

\begin{figure}
    \centering
    \includegraphics[width=\columnwidth]{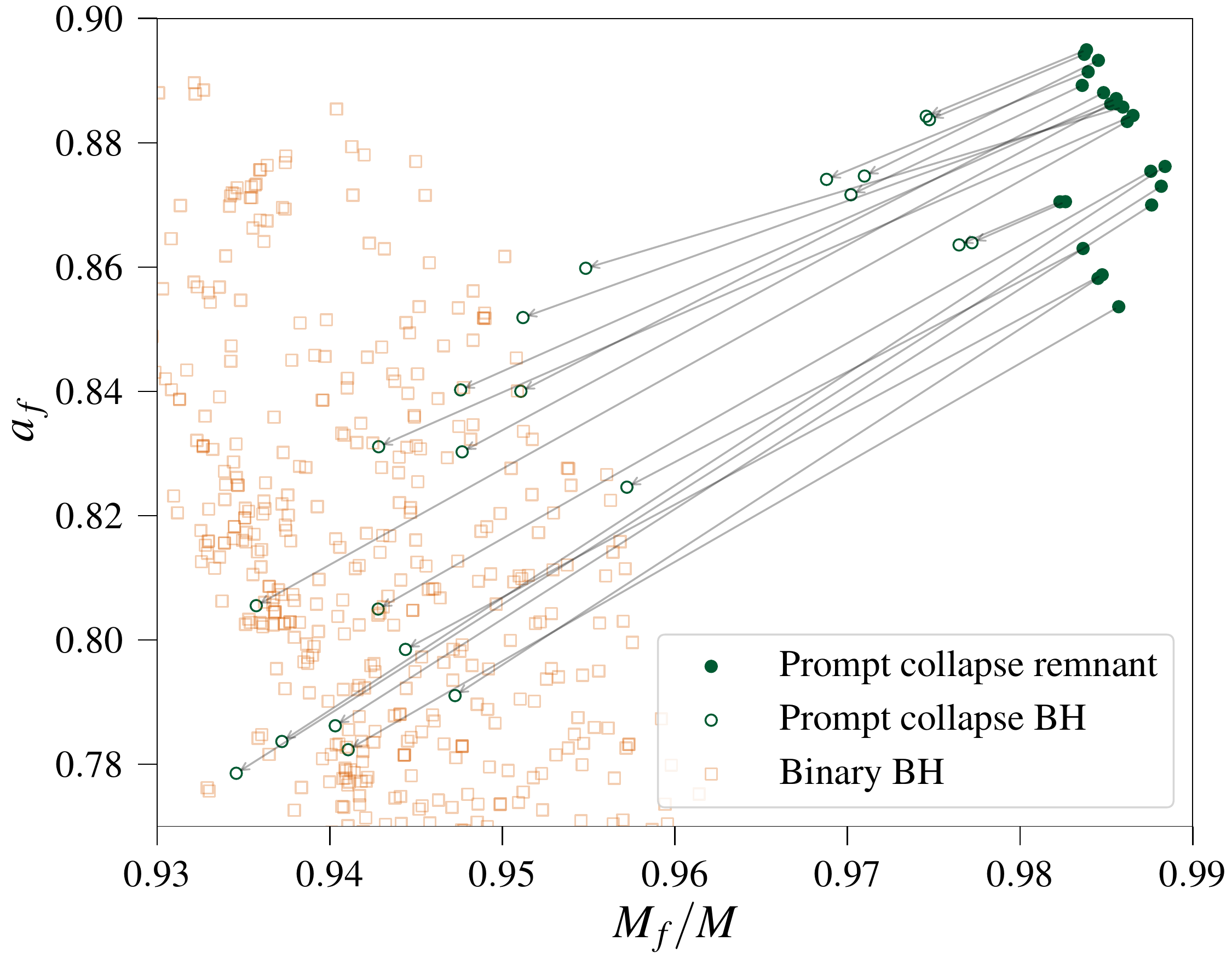}
    \caption{The dimensionless mass and spin of the remnant, including the disk (filled), and the remnant BH (unfilled; green) for the subset of 22 simulations for which we extract the accretion disk properties. The arrows depict the link between the remnant (including the disk) and the BH. We also show the remnant mass and spin of BBH mergers (unfilled; orange) for reference.}
    \label{fig:mf_af_disk_bh}
\end{figure}

We calculated the mass ($m_{disk}$) and the angular momentum ($J_{disk}$) contained in the disk outside the apparent horizon for a subset of 22 unequal mass simulations~\cite{Dhani:2025xno}. For these cases, the mass and dimensionless spin of the final BH can be estimated by
\begin{equation}
    M_f^{BH} = M_{adm} - E_{rad} - m_{disk},
\end{equation}
\begin{equation}
    a_f^{BH} = \frac{J_{adm} - J_{rad} - J_{disk}}{M_f^2}.
\end{equation}
We contrast the remnant properties with those of the remnant BHs in \cref{fig:mf_af_disk_bh}. We observe that there is a significant spread in the properties of the BHs, even though the remnants are tightly clustered. Indeed, the most asymmetric ones lie within the cluster of the BBH merger remnants. Nevertheless, it is worth remembering that the spacetime around the remnant BH is different from the BBH case due to the presence of an accretion disk. Note that some of the accreting matter will fall back into the BH on timescales of $\mathcal{O}(s)$ while the disk properties are extracted on timescales of $\mathcal{O}(ms)$. Therefore, at large timescales, the BH properties will be somewhere in between those depicted in \cref{fig:mf_af_disk_bh}.

\subsection{Implications of GW230529 on NS EoS}
\label{subsec:gw230529}
GW230529~\cite{LIGOScientific:2024elc} is the merger of a compact object with mass $3.6M_{\odot}$ and an NS of mass $1.4M_{\odot}$. The nature of the primary component is uncertain. It could be either the heaviest NS or the lightest BH observed to date. The formation channel for such an object is unclear as well. One of the possibilities is that it formed from a previous generation BNS merger~\cite{Mahapatra:2025agb}. Interestingly, it is spinning with a spin magnitude of $0.44^{+0.40}_{-0.37}$, and this allows us to make statements about the EoS of NS based on the discussion above.

Recall that $M_{\rm thr}$, the minimum total mass of a binary undergoing prompt collapse, is proportional to the maximum mass of a cold, non-rotating NS, i.e., $M_{\rm thr} = k_{\rm thr} M_{\max}$, $k_{\rm thr} \in [1.2, 1.6]$~\cite{Kashyap:2021wzs}. 
From \cref{fig:mf_af_disk_bh}, we can calculate the maximum total mass of the first-generation BNS merger---accounting for the energy released by the binary inspiral and merger---to be ${<}3.6/0.935\approx3.85\ M_{\odot}$. Note that we also assume that the accretion disk does not fall back into the BH following the first-generation merger. This gives the maximum threshold mass, $M_{\rm thr} \lesssim 3.85\ M_{\odot}$. If this first-generation merger underwent prompt collapse, that would imply $M_{\max} \lesssim 3.85/1.2 \approx 3.21 M_{\odot}$.
On the other hand, if the binary progenitor to the higher-mass object in GW230529 did not undergo prompt collapse, that would constrain the minimum maximum mass  to be $M_{\max} \gtrsim 3.85/1.6 \approx 2.41 M_{\odot}$. This latter scenario would be favored for small spins of the primary component of GW230529, in light of the results discussed in this work. This highlights the potential to use precise spin measurements to constrain the NS EoS.

\subsection{Detectability of the postmerger}
\label{subsec:detectability}
\begin{figure*}
    \centering
    \includegraphics[width=2\columnwidth]{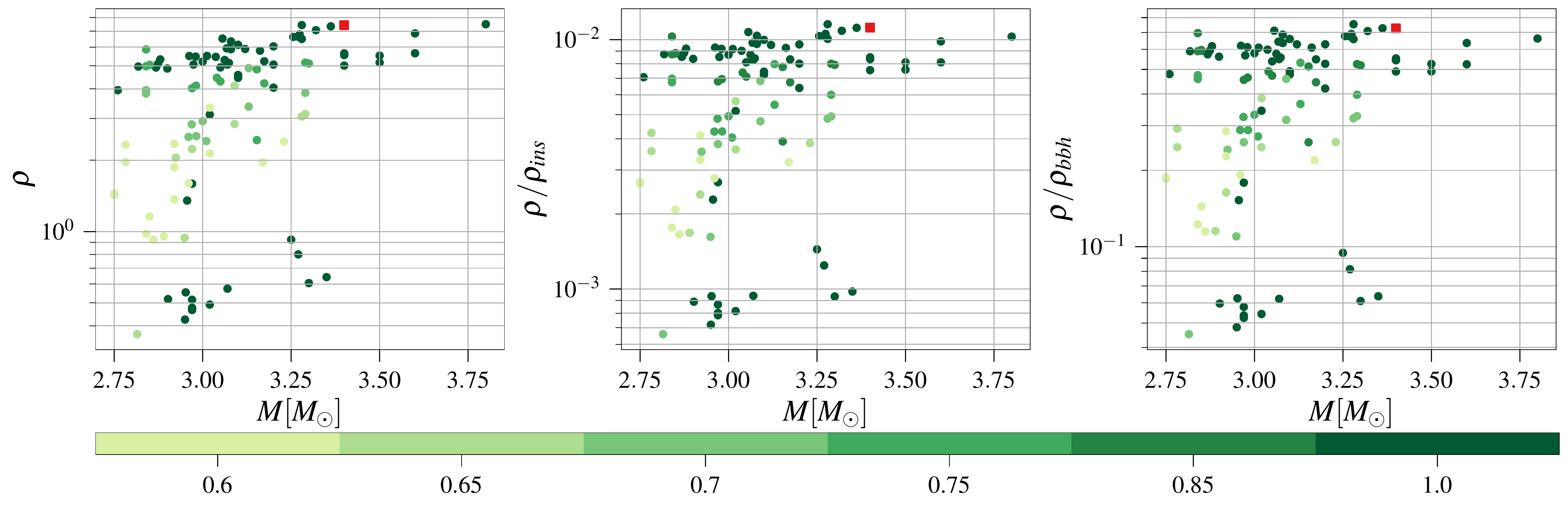}
    \caption{\emph{Left:} The SNR in the postmerger signal for the systems considered in this study placed at a distance of 100 Mpc and optimally oriented for a 40 km Cosmic Explorer detector. \emph{Center:} The SNR in the postmerger as a fraction of the inspiral SNR. \emph{Right:} The SNR in the postmerger as a fraction of an equivalent \ac{BBH} postmerger. As discussed in the main text, we separate the inspiral and postmerger signals using the cutoff frequency of 2048~Hz. The color gradient shows the asymmetry of the system with lighter shades representing more asymmetric binaries. Note that the colorbar is not uniform since simulations are only available at these fixed mass ratio values. The binary denoted with a red square is the $1.7\ M_\odot - 1.7\ M_\odot$ with the BA EoS, see the main text for further discussion.}
    \label{fig:snr}
\end{figure*}

It is instructive to calculate the expected SNR in the postmerger for the systems we consider. To this effect, we consider a 40 km Cosmic Explorer detector as a reference and place binaries at a distance of 100 Mpc. We expect 2-3 \ac{BNS} mergers from within a volume of radius 100 Mpc per year \cite{Borhanian:2022czq} based on the median rate of \ac{BNS} mergers from the second Gravitational-Wave Transient Catalog GWTC-2 \cite{LIGOScientific:2020ibl}. We assume that a binary is optimally located and oriented for the detector. The postmerger SNRs for such systems are then calculated using the formula
\begin{equation}
    \rho^2 = 4 \int_{f_{lo}}^{f_{hi}} \frac{|\tilde{h}(f)|^2}{S_h(f)} df,
\end{equation}
where $\tilde{h}(f)=F_+h_+ + F_{\times}h_{\times}$. $F_+$ and $F_{\times}$ are the antenna pattern functions of the detector, and $h_+$ and $h_{\times}$ are the two gravitational-wave polarizations that we have from numerical relativity simulations. In practice, for an optimal location in the sky with respect to the detector, we choose $F_+=1$ and $F_{\times}=0$. $S_h(f)$ is the noise power spectral density of the detector.

The postmerger SNR is very sensitive to the choice of the lower frequency since it lies at the heart of the sensitive band of the detector. We find that the merger frequency calculated directly from the NR data\footnote{The merger frequency is defined as the instantaneous frequency corresponding to the peak amplitude of the $\ell=2,m=2$ mode of the strain.} varied significantly from those predicted by the NR fits in \textcite{Dietrich:2020eud, Gonzalez:2022mgo}. The fits themselves varied considerably across a wide swath of the parameter space. Therefore, we make the choice to define the postmerger from a fixed $f_{lo} = 2048$ Hz. The high frequency cutoff is taken to be $f_{hi} = 7000$ Hz. 

In the left panel of Fig.~\ref{fig:snr}, we show the postmerger SNR for the binaries we consider. We find that most of them have SNRs greater than 4. These are also equal-mass or nearly equal mass binaries. Given that these are very short signals with extremely loud inspirals, even a SNR of 4 detection of such a signal will have a very small \emph{false alarm rate} and allow for reliable parameter estimation. For instance, GW150914 had a ringdown SNR of ${\sim}4.8$ (${\sim}8.5$) starting 6.5 ms (3 ms) after merger and a base-10-log Bayes factor $\log_{10}\mathcal{B}$ of ${\sim}3.5$ (${\sim}14$) for a damped sinusoid model of the ringdown against Gaussian noise \cite{LIGOScientific:2016lio}. The comparatively asymmetric binaries, as distinguished by their lighter shades in the figure, have SNRs between 0.9 and 4. However, we also find a select few (almost) equal mass binaries with a tiny post-merger SNR. These binaries ``shut-off'' immediately after their merger and there is no discernible post-merger radiation from them and hence have such small SNRs. 

In addition to the postmerger SNR, we calculate the SNR in the inspiral for all binaries where we take the starting lower frequency to be $f_{lo,insp} = 5$ Hz and the high frequency cutoff to be $f_{hi,insp} = 2048$ Hz.
We do not consider tidal effects in the amplitude of the inspiral waveform and, as a result, the SNR in the inspiral for an optimally located face-on system is given by,
\begin{equation}
    \rho^2 = 4 \frac{5}{24 \pi^{4/3}} \frac{\mathcal{M}_c^{5/3}}{D_L^2}\int_{f_{lo,insp}}^{f_{hi,insp}} \frac{f^{-7/3}}{S_h(f)} df,
\end{equation}
where $\mathcal{M}_c=(m_1 m_2)^{3/5}/(m_1+m_2)^{1/5}$ is the chirp mass of the binary.

We show the ratio of the SNR in the inspiral to the postmerger in the central panel of Fig.~\ref{fig:snr}. We find that for a majority of the systems considered, the postmerger has ${\sim}0.5{-}1\%$ of the inspiral SNR, and for a select few, it is much lower at $\lesssim0.1\%$.

Finally, we compare the postmerger SNR of the prompt collapse \ac{BNS} merger to the equivalent \ac{BBH} mergers. The SNR is calculated in the same frequency range of [2048, 7000] Hz as for the \ac{BNS} post-merger for an equivalent comparison. The \ac{BBH} waveform model is chosen to be \textsc{IMRPhenomD} \cite{Husa:2015iqa,Khan:2015jqa} which is a non-precessing waveform model containing only the $(l,|m|)=(2,2)$ mode.

We show the ratio of the SNRs in the prompt collapse ringdown to the \ac{BBH} ringdown in the right panel of Fig.~\ref{fig:snr}. We find that the SNR for \ac{BNS} is always smaller than the corresponding \ac{BBH} system. For a majority of the systems, the \ac{BNS} post-merger SNR is $40-80\%$ of the \ac{BBH} equivalent. However, for relatively asymmetric binaries, this can be even lower and in some cases contains only 10\% of their \ac{BBH} counterpart. Unsurprisingly, for systems with negligible post-merger, the fractional SNR is only a few percent. The smaller SNR for prompt collapse systems is expected from the previous section where it is shown that these systems are poorer GW emitters than \ac{BBH}s. The lack of power in the ringdown signal can be a smoking gun evidence for a \ac{BNS} system leading to a prompt collapse~\cite{Dhani:2023ijt}. We note that we have assumed the long-wavelength approximation for the detector response in our calculations. This needs to be relaxed in future studies.

Across the three panels, we find that the locations of the binaries are approximately constant, while the scale on the y-axis varies. This is because there is minimal variability in the BNS inspiral SNR or the BBH postmerger SNR across the different binary configurations.

\subsection{Distinguishing \ac{BNS} versus \ac{BBH} from postmerger}
\label{subsec:BNS_vs_BBH}
\begin{figure*}
    \centering
    \includegraphics[width=\columnwidth]{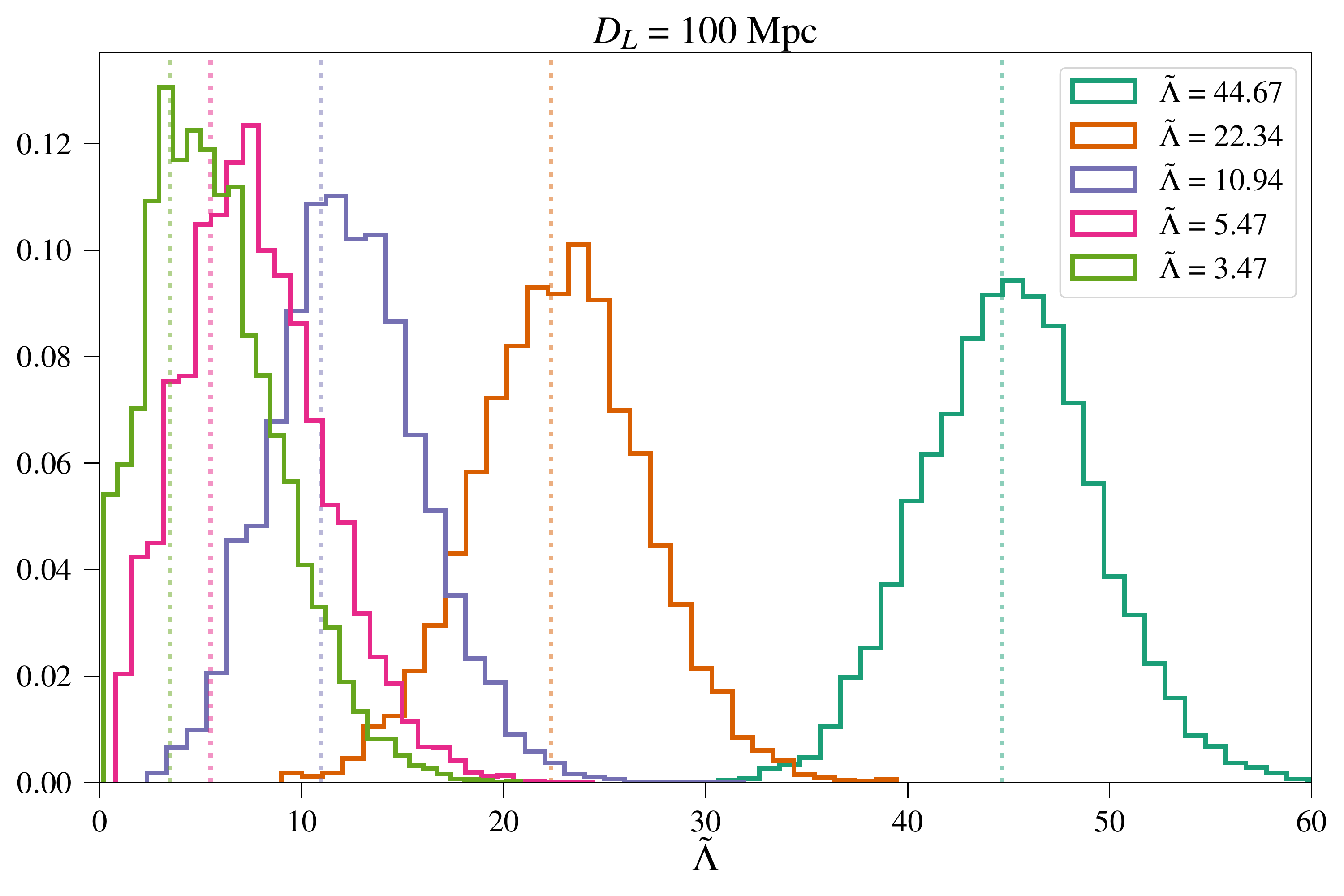}
    \includegraphics[width=\columnwidth]{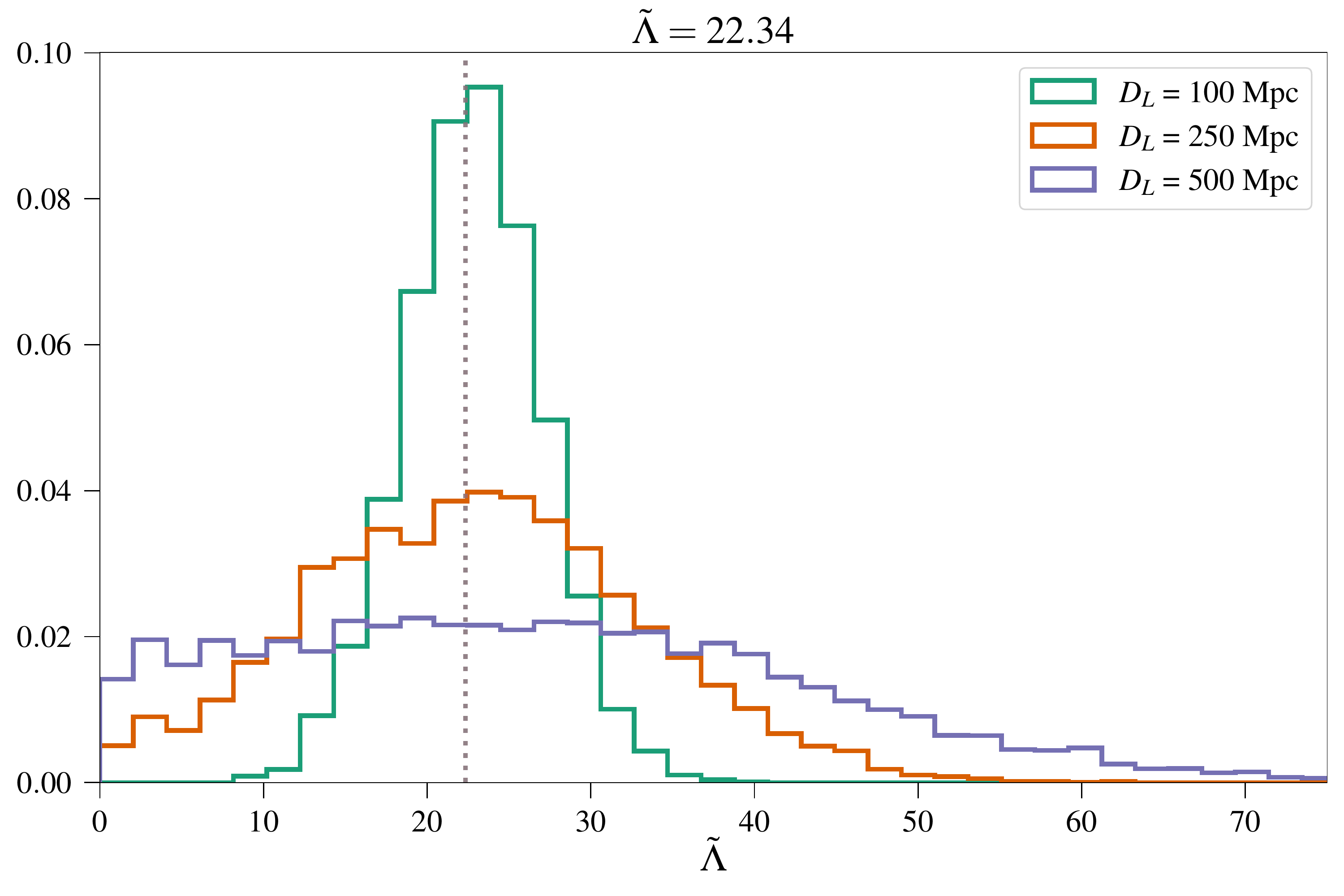}
    \caption{The probability density functions on the reduced tidal deformability parameter, $\tilde{\Lambda}$ with a 40 km Cosmic Explorer detector for a binary with redshifted masses (2.0 $M_{\odot}$, 1.9 $M_{\odot}$). \emph{Left:} The sources are kept at a fixed distance $D_L=100$ Mpc but the tidal deformabilities are varied. \emph{Right:} The reduced tidal deformability is kept constant at $\tilde{\Lambda}=22.34$ but the distance to the source is varied.}
    \label{fig:lamt}
\end{figure*}

In this section, we consider the distinguishability of a \ac{BNS} and \ac{BBH} merger using their postmerger signal if their inspiral is consistent with each other. To answer this, we first try to find the minimum tidal deformability parameter $\tilde{\Lambda}$ for which $\tilde{\Lambda}=0$ is excluded at a high confidence. We inject a set of high-mass non-spinning \ac{BNS} signals modeled using the \textsc{IMRPhenomPv2\_NRTidalv2}~\cite{Hannam:2013oca,Khan:2015jqa,Dietrich:2019kaq} waveform model in a single 40 km Cosmic Explorer detector. We choose the sky location and orientation of the binaries as that of GW170817 and the redshifted masses as (2.0 $M_{\odot}$, 1.9 $M_{\odot}$). We vary only the tidal deformabilities and the distance to the sources across the set. Our set of injected binaries can be divided into two non-mutually exclusive subsets. \emph{Set 1} contains 4 injections at $D_L=100$ Mpc with varying $\tilde{\Lambda} = \{44.67, 22.34, 10.94, 5.47, 3.47\}$. \emph{Set 2} contains 3 simulations with fixed $\tilde{\Lambda}=22.34$ at varying luminosity distances of $D_L = \{100, 250, 500\}$ Mpc. We note that these values for the tidal deformability do not correspond to any named EoS but were rather chosen by hand. We stress that this is inconsequential as long as the choices are in the ballpark of the EoS uncertainty. We use the publicly available GW simulation package \textsc{bilby} \cite{Ashton:2018jfp,Romero-Shaw:2020owr} to inject and recover the signal. 
We assume the binary's position in the sky is known since we are primarily interested in the tidal parameters, which are minimally correlated to the extrinsic parameters, such as the sky position.
We sample over all the other parameters of a non-spinning binary using the \textsc{PyMultinest} \cite{Buchner:2014nha} sampler and impose standard priors on all parameters, notably a uniform prior on the tidal deformability parameter $\tilde{\Lambda}$. 

The 1D marginalised posteriors on $\tilde{\Lambda}$ from parameter estimation across the two subsets of injections are shown in Fig.~\ref{fig:lamt}. The simulations corresponding to the binaries in Set 1 are shown in the left panel, while those corresponding to the binaries in Set 2 are depicted in the right panel. We find that a compact binary merger at $D_L=100$ Mpc can be confidently classified as a \ac{BNS} merger using its $\tilde{\Lambda}$ measurement for a value as small as $\tilde{\Lambda}\sim5$ and even $\tilde{\Lambda}\approx3.5$ can exclude 0 at $>$70\% confidence. Only a single EoS (SRO1) among the set of EoS we have considered attains a $\Lambda$ value smaller than 4. This is achieved for an NS mass of ${\sim}2.3 M_{\odot}$. An equal mass binary with a component mass of $2.3M_{\odot}$ would have a larger SNR than the masses considered in this analysis, so it would probably exclude 0 at an even higher confidence level. Therefore, up to a distance of $D_L=100$ Mpc, the inspiral signal can distinguish virtually all \ac{BNS} mergers from \ac{BBH} mergers. 

Next, we look at how the $\tilde{\Lambda}$ errors vary with distance. From the right panel of Fig.~\ref{fig:lamt}, we find that a $\tilde{\Lambda}\approx22$ can be measured to exclude the value 0 at a distance of $D_L=250$ Mpc. For larger distances, the $\tilde{\Lambda}$ posterior will exclude 0 at smaller confidence levels, as can be seen from the histogram of the same source placed at $D_L=500$ Mpc, and it will be difficult to conclusively identify such events as NS mergers. Moreover, at distances of $250$ Mpc and $500$ Mpc, the SNRs in the prompt collapse ringdowns will be smaller by factors of 2.5 and 5 as compared to the values shown in Fig.~\ref{fig:snr}. Neither \ac{BNS} nor \ac{BBH} ringdown will be clearly observable at such distances and hence a simple comparison of signal power in the ringdown cannot differentiate these systems. In such a scenario, accurate waveform templates describing the full signal is required to extract maximum information from the data and possibly distinguish them.

\section{Conclusion and Discussion}
\label{sec:conclusion}
The gravitational wave spectrum following a \ac{BNS} merger that does not collapse has a rich phenomenology that is very sensitive to the EoS used. Although there is a broad consensus on certain features of the post-merger spectrum, such a signal is not well modeled for faithful parameter estimation and reconstruction. 
On the other hand, prompt collapse BNS mergers contain complementary and largely EoS-agnostic information and can determine NS features such as the maximum mass supported by the NS EoS and the properties of the accretion disk around the remnant BH. 
To that end, we analyze 107 prompt collapse mergers across a set of EoSs, total mass, and mass ratios. 

We calculate the total energy radiated in gravitational waves and observe that it is very closely related to the angular momentum of the remnant. We verify that this relationship closely follows the relation proposed in \citet{Zappa:2017xba} and discuss its implications. 

We find that \ac{BNS} prompt collapses are poor GW emitters compared to their \ac{BBH} equivalent. 
We report that the spread in the scaled final mass and the dimensionless final spin across all the simulations considered is very narrow. This hints towards a minimal role played by the EoS in determining the dynamics of the very short prompt collapse timescale. We also observe that the final mass and spin of the remnant are larger than the product of an equivalent \ac{BBH} merger with the same initial configuration. In fact, the remnants are highly spinning with dimensionless spins between $0.85<a_f<0.95$. While for unequal mass systems, a portion of the mass and angular momentum is contained in the accretion disk surrounding the remnant BH, equal mass mergers have minimal matter accretion. Such systems can be of great astrophysical interest since it is otherwise difficult to produce highly spinning black holes. One can produce highly spinning BH remnants in a \ac{BBH} merger, but then one also has to start with relatively high spins for the progenitors, with the two values approaching each other the larger they become \cite{Kesden:2008ga, Hemberger:2013hsa}. 

We calculate the postmerger SNR of prompt-collapse \ac{BNS} mergers in the frequency band [2048, 7000] Hz and compare them to the signal strength of an equivalent \ac{BBH} post-merger. We report that the majority of the systems considered have post-merger SNRs $40-80\%$ relative to a \ac{BBH} post-merger with some as low as a few percent. We find that a majority of these systems would be observable up to a distance of 100 Mpc with a future 40 km CE. 

We examine the distinguishability of massive \ac{BNS} mergers with small tidal deformabilities that could result in prompt collapse from \ac{BBH} systems. We find that reduced tidal deformabilities as small as $\tilde\Lambda\approx3.5$ can be measured to exclude 0 up to a distance of 100 Mpc and $\tilde\Lambda\approx22$ to distances greater than 250 Mpc. 

Finally, we discuss the NS EoS implications of associating a prompt collapse to the progenitor of GW230529's primary component. We find that the maximum mass of NS is bound from above to ${\lesssim} 3.21 M_{\odot}$ if the primary is highly spinning, while it is constrained from below to ${\gtrsim} 2.41 M_{\odot}$ if it is not. These have implications on the maximum mass of NS \citep{Kashyap:2021wzs}.

Our work lays out the extremely interesting NS and GW physics in the 4~-~10~kHz region, which is often not the focus of technology and is only narrowly accessible to currently proposed GW detectors. 

A limitation of our study is that we use the long-wavelength approximation for the detector response for both SNR calculations and parameter estimation. We plan to relax this assumption in the future.
Another limitation of the study presented here is that only NR simulations with irrotational initial data are considered, although astrophysical \ac{BNS} are expected to be slowly spinning. We plan to extend the study to spinning \ac{BNS} and determine the maximal remnant BH spin that can be produced in a \ac{BNS} merger. It would also be interesting to understand the distribution of remnant properties for various configurations of black hole -- neutron star binaries. We leave that to the future as well.

\acknowledgements
We thank Aditya Vijaykumar for useful discussions. AD is supported by the NSF grant PHY-2012083.
DR acknowledges funding from the National Science Foundation under Grants No.~PHY-2020275, PHY-2116686, AST-2108467, PHY-2407681, the Sloan Foundation, and from the U.S. Department of Energy, Office of Science, Division of Nuclear Physics under Award Numbers DE-SC0021177 and DE-SC0024388.

\bibliography{BNS_ringdown}
\clearpage

\appendix
\section{Remnant mass -- spin}
\label{sec:appA}
\begin{figure}
    \centering
    \includegraphics[width=\columnwidth]{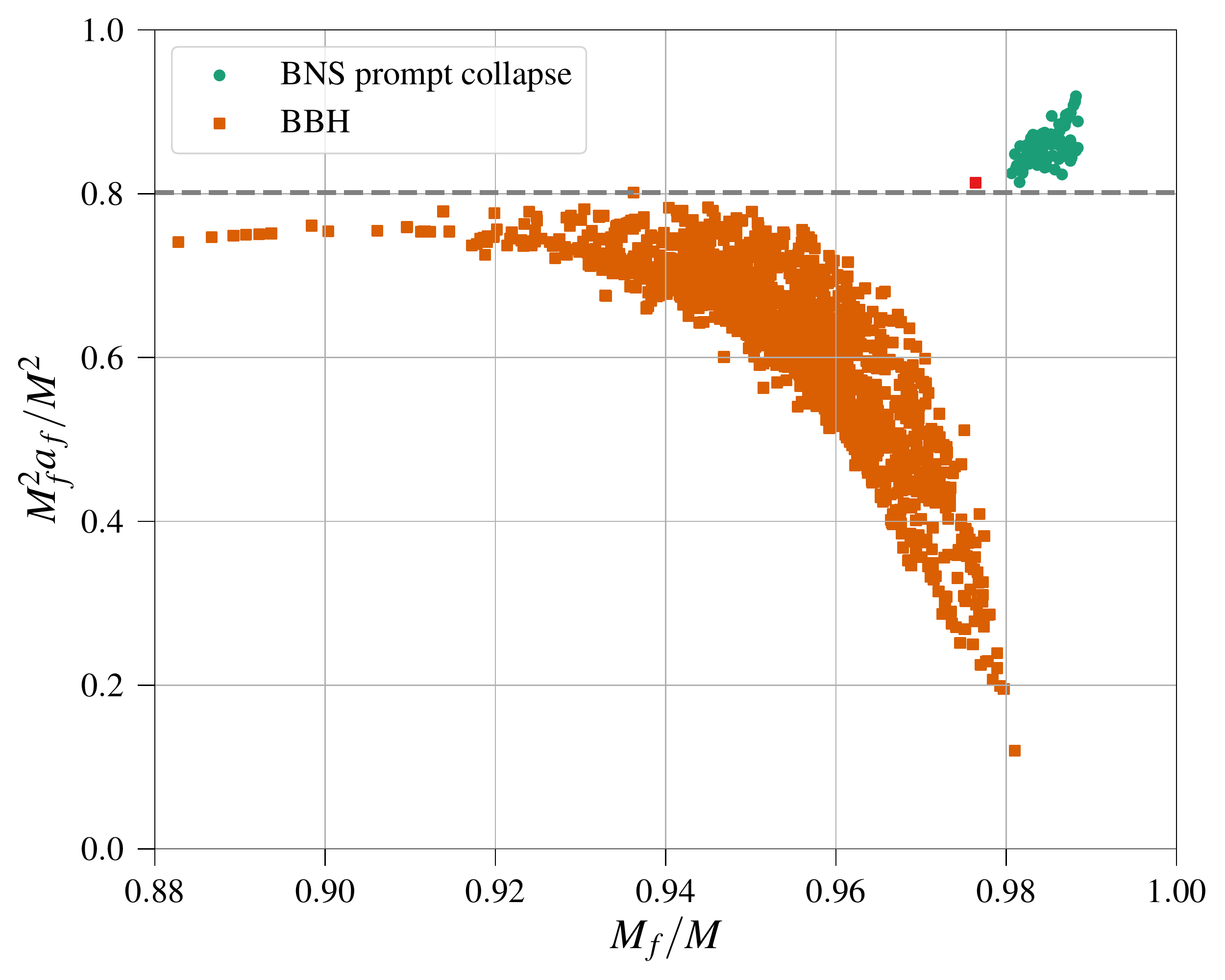}
    \caption{The remnant mass and angular momentum ($J_f/M^2$) of \ac{BBH} mergers versus \ac{BNS} prompt collapses. There are 1937 \ac{BBH} mergers in this figure corresponding to all the publicly available SXS simulations with $q \leq 3$ for which such information is available. The outlier denoted with a red square is the $1.7\ M_\odot - 1.7\ M_\odot$ with the BA EoS; see the main text for further discussion.}
    \label{fig:BBH_BNS_mf_spin}
\end{figure}

Fig.~\ref{fig:BBH_BNS_mf_spin} shows the angular momentum of the remnant scaled by the total mass of the binary for prompt collapses and \ac{BBH} mergers up to a mass ratio of 3 (see Sec.~\ref{subsec:remnant_mass_spin} for more details). It is observed that the angular momentum of the prompt collapse remnant per unit total mass is always greater than what is possible to achieve with \ac{BBH}. Note that the \ac{BBH} systems contain aligned spin and precessing binaries too.

\acrodef{BNS}{binary neutron star}
\acrodef{NS}{neutron star}
\acrodef{GW}{gravitational wave}
\acrodef{NR}{numerical relativity}
\acrodef{BH}{black hole}
\acrodef{QNM}{quasi-normal mode}
\acrodef{EM}{electromagnetic}
\acrodef{EoS}{equation of state}
\acrodefplural{EoS}[EoS]{equations of state}
\acrodef{CE}{Cosmic Explorer}
\acrodef{ET}{Einstein Telescope}
\acrodef{SNR}{signal-to-noise ratio}
\acrodef{BBH}{binary black holes}
\acrodef{EOB}{Effective one-body}
\acrodef{LWA}{long-wavelength approximation}
\acrodef{GRB}{gamma-ray burst}
\end{document}